\documentclass[preprint,12pt]{elsarticle}

\usepackage{amssymb}
\usepackage{amsmath}
\usepackage{graphicx}
\usepackage{subcaption}
\usepackage[normalem]{ulem}
\usepackage{tikz}
\usepackage{float}
\usepackage{multirow}
\usepackage{hyperref}
\usepackage{xcolor}

\journal{Chemical Physics}

\begin{document}

\begin{frontmatter}



\title{Rotational laser spectroscopy of the $X^1\Sigma_g^+\rightarrow B^1\Pi_u$ transition of $\text{Rb}_2$ molecule in a supersonic beam: As good as it gets} %% Article title


\author[label1]{David Rodríguez Fernández} %% Author name

\author[label1]{Manuel Alejandro Lefrán Torres} %% Author name

\author[label1]{Jaime Javier Borges Márquez} %% Author name

\author[label1]{Marcos Roberto Cardoso} %% Author name

\author[label2]{Amrendra Pandey} %% Author name

\author[label2]{Romain Vexiau} %% Author name

\author[label2]{Olivier Dulieu} %% Author name

\author[label2]{Nadia Bouloufa-Maafa} %% Author name

\author[label1]{Luis Gustavo Marcassa} %% Author name

\affiliation[label1]{organization={Instituto de Física de São Carlos, Universidade de São Paulo},%Department and Organization
            addressline={Avenida Trabalhador São-carlense, nº 400}, 
            city={São Carlos},
            postcode={13566-590}, 
            state={São Paulo},
            country={Brasil}}

\affiliation[label2]{organization={Université Paris-Saclay, CNRS, Laboratoire Aimé Cotton},%Department and Organization
            city={Orsay},
            postcode={91400}, 
            country={France}}            

%% Abstract

\begin{abstract}
High-resolution laser spectroscopy of the $^{85}\mathrm{Rb}_2$, $^{85}\mathrm{Rb}^{87}\mathrm{Rb}$, and $^{87}\mathrm{Rb}_2$ isotopologues has been performed in a supersonic molecular beam using a continuous-wave (cw) tunable diode laser. A total of 958 rovibronic transitions were recorded up to $7~\mathrm{cm}^{-1}$ below the vibrational band heads of the $X^1\Sigma_g^+(v''=0)\rightarrow B^1\Pi_u(v'=1,2)$ and $X^1\Sigma_g^+(v''=1)\rightarrow B^1\Pi_u(v'=1)$ bands, with a spectral resolution of $3.3\times10^{-4}$~cm$^{-1}$. Although restricted to the $v'=1$ and $v'=2$ vibrational levels of the excited $B^1\Pi_u$ state, the measurements extend previous work by Amiot and Vergès [\textbf{Chemical Physics Letters 274, 91 (1997)}] through substantially higher resolution and dense low-$J'$ rotational data for all three isotopologues. A global least-squares analysis combining the new and published data, yields improved Dunham coefficients for the excited $B^1\Pi_u$ state, significantly refines rotational and rovibrational coupling constants. In addition, the $\Lambda$-doubling constants of $B^1\Pi_u$ state were determined for the three isotopologues.

\end{abstract}

\begin{graphicalabstract}
\begin{figure}[h]
\centering
\includegraphics[width=1\textwidth]{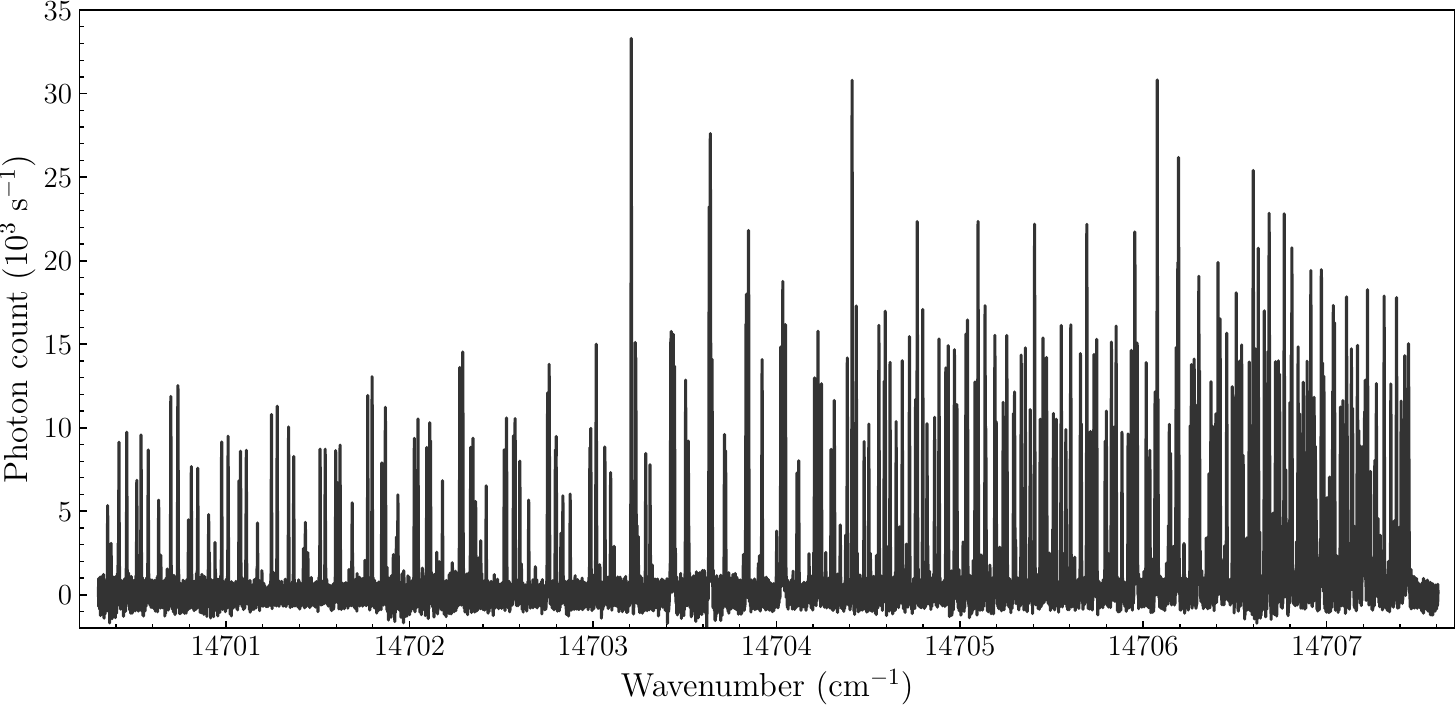}
\end{figure}
\end{graphicalabstract}

\begin{highlights}
\item The rovibrational lines corresponding to the isotopologues  $^{85}\mathrm{Rb}_2$,  $^{85}\mathrm{Rb}^{87}\mathrm{Rb}$, and  $^{87}\mathrm{Rb}_2$ are clearly resolved and identified for the $X^1\Sigma_g^+(v''=0)\rightarrow B^1\Pi_u (v'=1,2)$ vibrational bands.

\item The lines for the $X^1\Sigma_g^+(v''=1)\rightarrow B^1\Pi_u (v'=1)$ vibrational band are identified only for $^{85}\mathrm{Rb}_2$ and $^{85}\mathrm{Rb}^{87}\mathrm{Rb}$.

\item The $\Lambda$-doubling of the $B^1\Pi_u$ state of the $\mathrm{Rb}_2$ molecule is measured.

\item A global fitting is implemented, considering the present data and those from C. Amiot and J. Vergès.
\end{highlights}

%% Keywords

\end{frontmatter}



\section{Introduction}
\label{sec1}

Since their invention in the mid 20th century \cite{ramsey1956}, molecular beams represent a technique of choice for accurate molecular spectroscopy. Despite their high reactivity, the complex spectrum of alkali-metal diatomic molecules have been addressed by many groups, as illustrated for example in the book of W.C. Stwalley and collaborators which includes a huge list of references impossible to reproduce here \cite{kim2014}. With the advances of laser cooling and trapping in the 90' which has been pioneered with alkali-metal species, numerous breakthrough have been achieved in the formation of ultracold diatomic molecules, either homonuclear or heteronuclear, which has allowed complementary analysis of their molecular spectra, for instance via the so-called photoassociative spectroscopy \cite{stwalley1999,stwalley2012,pazyuk2015}. 

Merging molecular spectroscopic data is a must to overcome the accuracy limitation inherent to a given technique. This complementarity has been beautifully demonstrated on several excited electronic states of Rb$_2$ \cite{amiot1995,salami2009} and Cs$_2$ \cite{bai2011}, which knowledge has been improved by merging molecular data from various sources, including laser-induced fluorescence, Fourier transform spectroscopy, optical-optical double resonance polarization spectroscopy, and photoassociative spectroscopy.

The excited electronic state $B^1\Pi_u$ of Rb$_2$ correlated to the dissociation limit Rb($5s$)+Rb($5p$) plays a major role in spectroscopic investigations. Its excitation serves as an intermediate step in the study of high excited electronic states \cite{PASHOV2022,PhysRevA.105.032823,PhysRevA.109.012805,Drozdova2012}. Lefrán and co-workers have proposed using the state $B^1\Pi_u$ for laser cooling of Rb$_2$ molecules \cite{Alejandro2023}. The ionization energy of the Rb$_2$ molecule was measured via resonantly enhanced two-photon ionization (RE2PI) in a supersonic beam through the state $B^1\Pi_u$ \cite{Lefran2025}. This RE2PI process through the  $B^1\Pi_u$ state was recently used to probe ultracold chemistry between KRb molecules and quantify their products \cite{Hu2021,Liu2021}.

Surprisingly, to our knowledge, there are only two spectroscopic investigations of the state $B^1\Pi_u$ of Rb$_2$. In the early 1980's, Caldwell \textit{et al.} performed laser induced fluorescence in a supersonic beam \cite{Caldwell1980}, while a more extensive work was carried out by C. Amiot and J. Vergès using Fourier transform spectroscopy in a heat pipe \cite{Amiot1997}. However, the knowledge of the lowest vibrational levels of this state remains limited, as noted in \cite{Drozdova2012, drozdova2014}. In these works, the spectroscopy of the high-lying state $2^1\Pi_g$ has been reported using double-resonance excitation, and the fluorescence toward the state $B^1\Pi_u$ has been analyzed, are claimed to yield a better representation of its lower vibrational levels.

In the present paper, we report high-resolution laser spectroscopic measurements of $\mathrm{Rb}_2$ molecules generated in a supersonic molecular beam. By tuning a diode laser, we drive rotational transitions in the $X^1\Sigma_g^+(v''=0)\rightarrow B^1\Pi_u (v'=1,2)$ and $X^1 \Sigma_g^+(v''=1)\rightarrow B^1\Pi_u (v'=1)$ vibrational bands. The rovibrational features associated with the isotopologues $^{85}\mathrm{Rb}_2$, $^{85}\mathrm{Rb}^{87}\mathrm{Rb}$, and $^{87}\mathrm{Rb}_2$ are distinctly resolved and assigned. By merging our data with those of \cite{Amiot1997}, we performed a global fit that extends and refines earlier works \cite{Caldwell1980, Amiot1997}, significantly improving spectral resolution. We also determine the $\Lambda$-doubling of the $B^1\Pi_u$ state of the $\mathrm{Rb}_2$ molecule. 

\section{Experimental setup}
\label{sec2}

We adopted an experimental approach similar to that
of Caldwell \textit{et al.} \cite{Caldwell1980}. Briefly, the $\mathrm{Rb}_2$ molecules are created by a supersonic oven, operating at a temperature of $873.15~\text{K}$ and a pressure of $\approx 368~\text{torr}$ \cite{SteckRb85}.  The nozzle, with a diameter of $300~\mu\text{m}$, is kept at a temperature 50 K above the oven temperature to avoid clogging. The molecular beam expands into a vacuum chamber maintained at a pressure of $10^{-5}~\text{torr}$, evacuated by a diffusion pump, supported by a mechanical pump. A more detailed description of the apparatus is provided in \cite{Lefran2025}. The molecular beam is collimated by a skimmer, with a diameter of $3~\mathrm{mm}$, placed $26~\mathrm{mm}$ from the nozzle exit, resulting in a divergence angle of approximately $115~\mathrm{mrad}$.

\begin{figure}[H]
\centering
\includegraphics[width=1.0\textwidth]{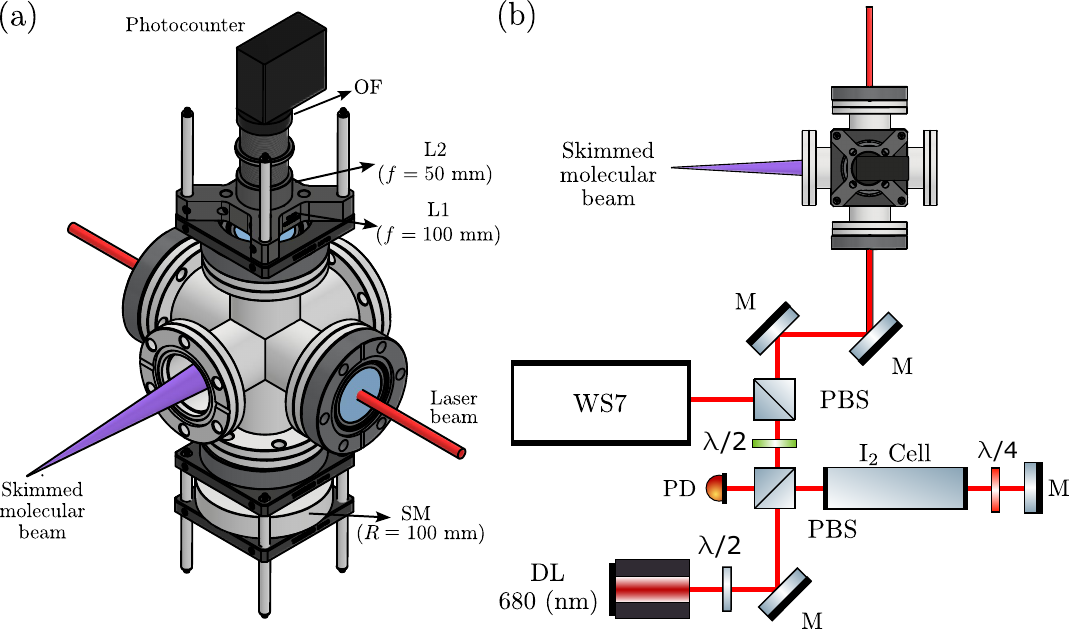}
\caption{Experimental setup for fluorescence spectroscopy of $\text{Rb}_2$ molecules produced in a supersonic beam. (a) Schematic of the fluorescence detection chamber. The molecular beam, generated from a heated rubidium oven and collimated by a skimmer, is intersected perpendicularly by the excitation laser in the detection region. Fluorescence photons are collected through two opposite optical windows. The upper fluorescence signal is collimated by a lens (L1) and focused by a second lens (L2) onto a photon-counting module, while the fluorescence emitted toward the lower window is retro-reflected by a spherical mirror to enhance the collection efficiency. A band-pass optical filter (OF) is used to suppress background light. (b) Schematic of the laser excitation and frequency calibration system. The diode laser (Toptica DL pro) operates at a 680 nm wavelength. The laser frequency is monitored by a wavelength meter (HighFinesse WS7) and an iodine cell. The laser beam intersects the molecular beam in the detection region, enabling the acquisition of fluorescence spectra by frequency scanning.
}
\label{fig:1}
\end{figure}

After passing through the diffusion pump chamber, the skimmed molecular beam enters the fluorescence detection chamber, whose pressure is $2\times 10^{-6}~\text{torr}$. The detection chamber consists of a six-way cross, as shown in Figure \ref{fig:1}(a), with two opposing side arms fitted with antireflection-coated optical windows that provide access to the excitation laser. Fluorescence spectroscopy was performed using a diode laser (Toptica DL Pro laser) operating in the range $14640-14760~\mathrm{cm}^{-1}$ with a linewidth of 500 kHz (FWHM). The laser frequency was continuously monitored using a wavelength meter (HighFinesse WS7), combined with an iodine cell \cite{LefrnTorres2022, Fernndez2023}. The laser beam intersects the skimmed molecular beam perpendicularly in the detection region, located 16.3 cm downstream from the nozzle, as shown in Figure (\ref{fig:1}(a)). Spontaneously emitted photons are collected through the upper and lower optical windows. The light emerging from the upper window is collimated by a spherical lens (L1) with focal length $f = 100~\mathrm{mm}$ and subsequently focused by a lens (L2) with $f = 50~\mathrm{mm}$ onto a Hamamatsu H11890-01 photon-counting module. Fluorescence emitted toward the lower window is retro-reflected by a spherical mirror with radius of curvature $R = 100~\mathrm{mm}$ and recombined with the upper optical path to enhance collection efficiency. To suppress background noise, an optical band-pass filter (OF) (Thorlabs FBH680-10) was placed in front of the detector. By scanning the laser frequency and recording the photon-counting signal, fluorescence spectra are obtained. 

Under the present experimental conditions, the vibrational population of the molecular beam is predominantly concentrated in the lowest vibrational levels, namely $v''=0$ and $v''=1$. This population distribution allows us to investigate in detail the rotational structure of the vibrational bands $X^1\Sigma_g^+(v''=0)\rightarrow B^1\Pi_u(v'=1,2)$ and $X^1\Sigma_g^+(v''=1)\rightarrow B^1\Pi_u(v'=1)$.

\section{High-Resolution Rotational Fluorescence Spectra  and Line Assignment} \label{sec3} 

\begin{figure}[h]
\centering
\includegraphics[width=\textwidth]{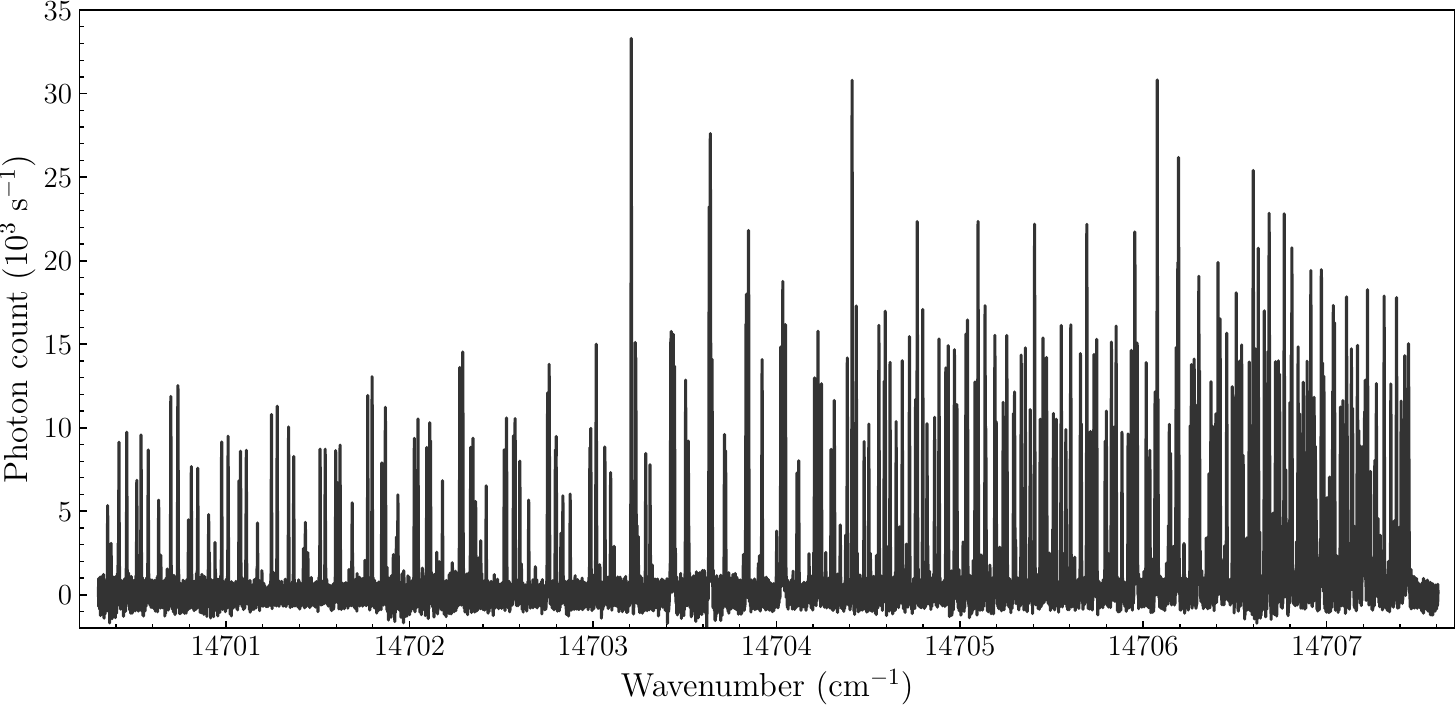}
\caption{Fluorescence spectrum of $\mathrm{Rb}_2$ recorded over a $\sim 7~\mathrm{cm}^{-1}$ range below the $v''=0 \rightarrow v'=1$ vibrational band head. The spectral resolution is $3.3\times10^{-4}~\mathrm{cm}^{-1}$.}
\label{fig:2}
\end{figure}

We have recorded fluorescence spectra over three energy intervals, $(14700.3-14707.6)~\mathrm{cm}^{-1}$, $(14747.0-14754.4)~\mathrm{cm}^{-1}$, and $(14643.0-14650.4)~\mathrm{cm}^{-1}$, corresponding to regions located approximately $7~\mathrm{cm}^{-1}$ below the vibrational band heads of the $v'' = 0 \rightarrow v' = 1$, $v'' = 0 \rightarrow v' = 2$, and $v'' = 1 \rightarrow v' = 1$ transitions, respectively. Figure~\ref{fig:2} presents the spectrum recorded below the vibrational band head of the $v'' = 0 \rightarrow v' = 1$ transition. The two other spectra analyzed in this work are provided in the Supplementary Material \cite{supp}.

The spectra are recorded with a step resolution of 10 MHz. They exhibit a satisfactory signal-to-noise ratio $(\sim16~\mathrm{dB})$, allowing a reliable identification of the rotational structure of the three isotopologues $^{85}\mathrm{Rb}_2$, $^{85}\mathrm{Rb}^{87}\mathrm{Rb}$, and $^{87}\mathrm{Rb}_2$. The transition linewidth (FWHM), extracted from the peaks, is approximately 60~MHz, which is consistent with the Doppler broadening arising from the velocity distribution of the molecular beam imposed by the nozzle-skimmer geometry of our experimental setup. The broadening mechanisms associated with the second-order Doppler effect, spontaneous emission, and power saturation are negligible under the present experimental conditions.

To determine the spectral positions of each fluorescence peak, we applied the same automated peak-detection procedure used in a recent $I_{2}$ work~\cite{LefrnTorres2022, Fernndez2023}. The \texttt{find\_peaks} function from the Python SciPy library \cite{Virtanen2020} was used to locate local maxima by comparing them with neighboring data points. The height and distance parameters of the function \texttt{find\_peaks} were chosen according to physically motivated criteria: the height threshold was set above the noise level, and the minimum distance between adjacent peaks was constrained to be equal to or greater than the peak linewidth. 

The peaks associated with the rotational transitions were identified and assigned to the corresponding isotopologues by applying the theory of electronic optical transitions in diatomic molecules together with the procedure presented in the following. The Dunham expansion of the energies $E''^{(i)}(v'',J'')$ of the rovibrational levels $(v'',J'')$ of the ground state $X^1\Sigma_g^+$ of a given isotopologue $(i)$ ($i=\{1,2,~\mathrm{and}~3\}$ corresponds to $^{85}\mathrm{Rb}_2$, $^{85}\mathrm{Rb}^{87}\mathrm{Rb}$, and $^{87}\mathrm{Rb}_2$, respectively) can be written as

\begin{align}
 E''^{(i)}(v'',J'')=hc\sum_{k,l} Y_{k,l}''\, \rho_i^{\frac{k}{2}+l}\, \left(v''+\tfrac{1}{2}\right)^k X''^l, 
 \label{eq:1}
\end{align}
where $X''=J''(J''+1)$ and $\rho_i = \mu/\mu_i$ is the ratio of the reduced mass $\mu$ of the reference isotopologue $^{85}\mathrm{Rb}_2$ to the reduced mass of the isotopologue indexed by $i$ \cite{DeLaeter2003}. The terms $Y''_{k,l}$ denote the Dunham coefficients of the reference isotopologue. The term $Y''_{10}$ is the harmonic vibrational constant  $\omega_e''$, and $Y''_{20}$ is the leading anharmonic contribution usually expressed as $-(\omega_e x_e)''$. The coefficients $Y''_{30}$, $Y''_{40}$, etc. account for higher order anharmonic corrections. The term $Y''_{01}$ is the rotational constant $B_e$ of the state and the terms $l>1$ account for high-order anharmonic corrections.

The Dunham expansion for $B^1\Pi_u$ state contains additional terms due to the $\Lambda$-doubling which induces parity splitting of $+$ and $-$ states,

\begin{multline}
E'^{(i)}(v',J',\pm)= hc\sum_{k}\left(v'+\frac{1}{2}\right)^k
\Bigg\{\sum_lY_{k,l}'\rho_i^{\,k/2+l}X'^l
\pm\frac{1}{2}
\left[p_k\rho_i^{1/2} X'+q_k\rho_i X'^2\right]\Bigg\},
\label{eq:2}
\end{multline}
where $X'=J'(J'+1)-1$. The $\Lambda$-doubling constants $p_k$ and $q_k$ exhibit a vibrational dependence and are expanded as a power series in ($v'+1/2$). However, since only the $v'=1$ and $v'=2$ vibrational levels are available in the present analysis, this dependence can not be reliably determined. Therefore, only the zeroth-order constants $p_0$ and $q_0$ were kept in the fit yielding the simplified Dunham expansion
\begin{equation}
\begin{aligned}
E'^{(i)}(v',J',\pm)=hc\sum_{k,l}Y'_{kl}\,
\rho_i^{\,k/2+l}
\left(v'+\frac12\right)^k
X'^l
&\pm\frac12
\Bigg[
p_0\,\rho_i^{1/2}X'+q_0\,\rho_iX'^2\Bigg].
\label{eq:3}
\end{aligned} 
\end{equation}
Then, the transition wavenumber between the excited and ground states is given by 
\begin{align}
\nu^{(i)}(v'',J'',v',J',\pm) = (E'^{(i)}(v', J', \pm)-E''^{(i)}(v'', J''))/hc,
\label{eq:4}
\end{align}
where both the Dunham coefficients and the $\Lambda$-doubling constants are expressed in $\mathrm{cm}^{-1}$.
The difference $hc(Y'_{00}-Y''_{00})$ is the energy gap between the minima of the potential energy curves of the initial and final electronic states. The electronic transition $X^{1}\Sigma_g^+ \rightarrow B^{1}\Pi_u$ gives rise to the usual rotational branches $R$, $Q$, and $P$ for $J'-J''=+1,0,-1$, respectively. As a first step in the line assignment procedure, the transition energies for a given isotopologue were calculated to the second order while neglecting the $\Lambda$-doubling. The resulting expressions for the $R$, $Q$, and $P$ branches are, respectively: 
\begin{align}
\nu_R^{(i)}(v'',v', J'') &= \nu^{(i)}(v'',v') + Y'^{(i)}_{01} [(J''+2)(J''+1)-1] + Y'^{(i)}_{02} [(J''+2)(J''+1)-1]^2 +...\notag \\
 & - Y''^{(i)}_{01} J''(J''+1) - Y''^{(i)}_{02} [J''(J''+1)]^2,  \label{eq:5}\\
 \nu_Q^{(i)}(v'',v',J'') &= \nu^{(i)}(v'',v') + Y'^{(i)}_{01}[J''(J''+1)-1] + Y'^{(i)}_{02} [J''(J''+1)-1]^2+ ... \notag \\ 
& - Y''^{(i)}_{01} J''(J''+1) - Y''^{(i)}_{02} [J''(J''+1)]^2, \label{eq:6}\\
 \nu_P^{(i)}(v'',v',J'') &= \nu^{(i)}(v'',v') + Y'^{(i)}_{01}[J''(J''-1)-1] + Y'^{(i)}_{02} [J''(J''-1)-1]^2 \notag+ ...
 \\ 
& - Y''^{(i)}_{01} J''(J''+1) - Y''^{(i)}_{02} [J''(J''+1)]^2, 
 \label{eq:7}
\end{align}
where the origin of the vibrational band $v''\rightarrow v'$ is given by
\begin{align}
\nu^{(i)}(v'',v') = \sum_k\rho_i^{k/2}\{Y'_{k,0}(v'+1/2)^k - Y''_{k,0}(v''+1/2)^k\}.
 \label{eq:8}
\end{align}
%  

%For the assignment of the rotational transitions, our analysis was initially restricted to a $0.15~\mathrm{cm}^{-1}$ interval below the band head, shown in Figure \ref{fig:3}(a). Since the rotational constant of the electronic ground state is greater than that of the excited state $(Y''_{01} > Y'_{01})$ \cite{Seto2000,Amiot1997}, the expected spectral behavior is the formation of a $R$ branch band head. In this regime, the transition frequencies of the $R$ branch initially increase with rotational quantum number, reach a maximum at the band head, leading to characteristic line crowding in this region, and subsequently decrease with a further increase of $J''$.}

\begin{figure}[H]
\centering
\includegraphics[width=0.75\textwidth]{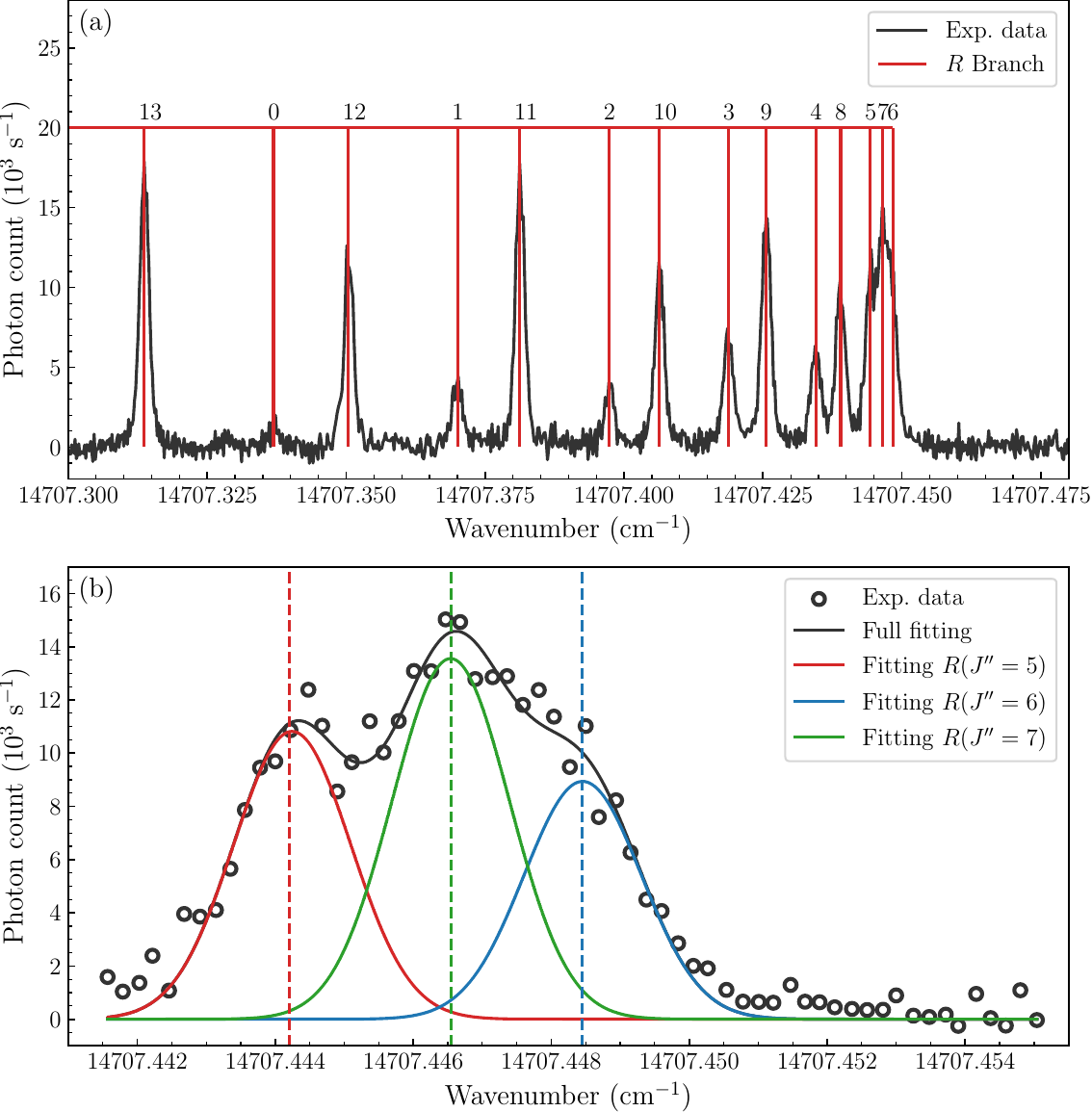}
\caption{Rotationally resolved spectrum near the band head of the 
transition $X^1\Sigma_g^+(v''=0) \rightarrow B^1\Pi_u(v'=1)$ in $^{85}\mathrm{Rb}_2$. (a) Expanded view of a $0.15~\mathrm{cm}^{-1}$ spectral range below the band head, showing the assignment of the $R$-branch rotational lines up to the $R(J''=13)$. (b) Enlarged view of the band head region, where three partially unresolved rotational transitions are observed. The solid curve corresponds to the best fit using a three-component Gaussian profile.}
\label{fig:3}
\end{figure}

For the assignment of the rotational transitions, our analysis was initially limited to a $0.15~\mathrm{cm}^{-1}$ spectral interval below the band head, as shown in Figure \ref{fig:3}(a). Since the rotational constant of the state $X^1\Sigma_g^+$ ($Y''_{01} \approx 2.2404336\times10^{-2} ~\mathrm{cm}^{-1}$ \cite{Seto2000}) is greater than the one of the state $B^1\Pi_u$ ($Y'_{01} \approx 1.9546311\times10^{-2} ~\mathrm{cm}^{-1}$ \cite{Amiot1997}), the formation of a $R$-branch band head is expected. In this region, the $R$-branch transition energies initially increase with increasing rotational quantum number, reach a maximum corresponding to the band head position, and then decrease as $J''$ continues to increase. This behavior results in a characteristic accumulation of rotational lines near the band head.

Line assignment was carried out by comparison with transition energies calculated using the well-established Dunham coefficients of the state $X^1\Sigma_g^+$\cite{Seto2000}, which were kept fixed throughout the analysis, together with the excited-state Dunham coefficients reported in \cite{Amiot1997}. The calculated energy separation between the $R(J''=0)$ transition and the band head for $^{85}\mathrm{Rb}_2$ was used to identify the $R(J''=0)$ line, observed at $14707.3373~\mathrm{cm}^{-1}$. The positions of the subsequent rotational lines were then predicted from the rotational constants and compared with the experimental spectrum. The excellent agreement obtained for the next 13 rotational lines confirmed the rotational assignment. The relative intensities of neighboring odd- and even-$J$ rotational lines, governed by the nuclear-spin statistical weights \cite{Aldegunde2009}, provided additional support for the assignment. In the band head region, where the rotational structure becomes partially unresolved, the central positions of the individual components were determined by means of a three-component Gaussian fit (Figure~\ref{fig:3}(b)).

%Since the Dunham coefficients of the ground $\Sigma$ state are well established \cite{Seto2000}, they were kept fixed in our  analysis. Using the rotational constants $Y'_{01}$ and $Y'_{02}$ reported in \cite{Amiot1997}, we calculated the energy separation between the $R(J''=0))$ transition and the band head for $^{85}\mathrm{Rb}_2$ . The relative intensities of neighboring odd and even rotational lines, governed by the nuclear-spin statistical weights \cite{Aldegunde2009}, were also used as an additional indication to support the rotational assignment. Firstly, The transition at $14707.3373~\mathrm{cm}^{-1}$  was assigned to $R(J''=0)$ then using the rotational constants from \cite{Amiot1997}, the positions of the following 13 rotational lines were predicted and found to be in excellent agreement with the experimentally observed peak positions. In the band head region, where rotational lines become partially unresolved, a three-component Gaussian fit was performed to determine the central positions of the individual peaks $Figure~\ref{fig:3}(b)$. }

The assigned transitions were then used to refine the fit parameters and predict additional rotational lines (Eq.~\ref{eq:5}). This procedure was repeated several times, gradually extending the range of assigned transitions until all lines of the $R$ branch were identified. The final set of parameters was later used to assign the $Q$ and $P$ branch transitions using Eqs.~(\ref{eq:6}) and (\ref{eq:7}). It should be noted that for high rotational quantum numbers, the observed $Q$-branch transitions exhibit slight deviations from the simple theoretical model, as discussed later. Nevertheless, the agreement remains sufficiently good to ensure reliable line assignments over the investigated rotational range.

\begin{figure}[H]
\centering
\includegraphics[width=0.8\textwidth]{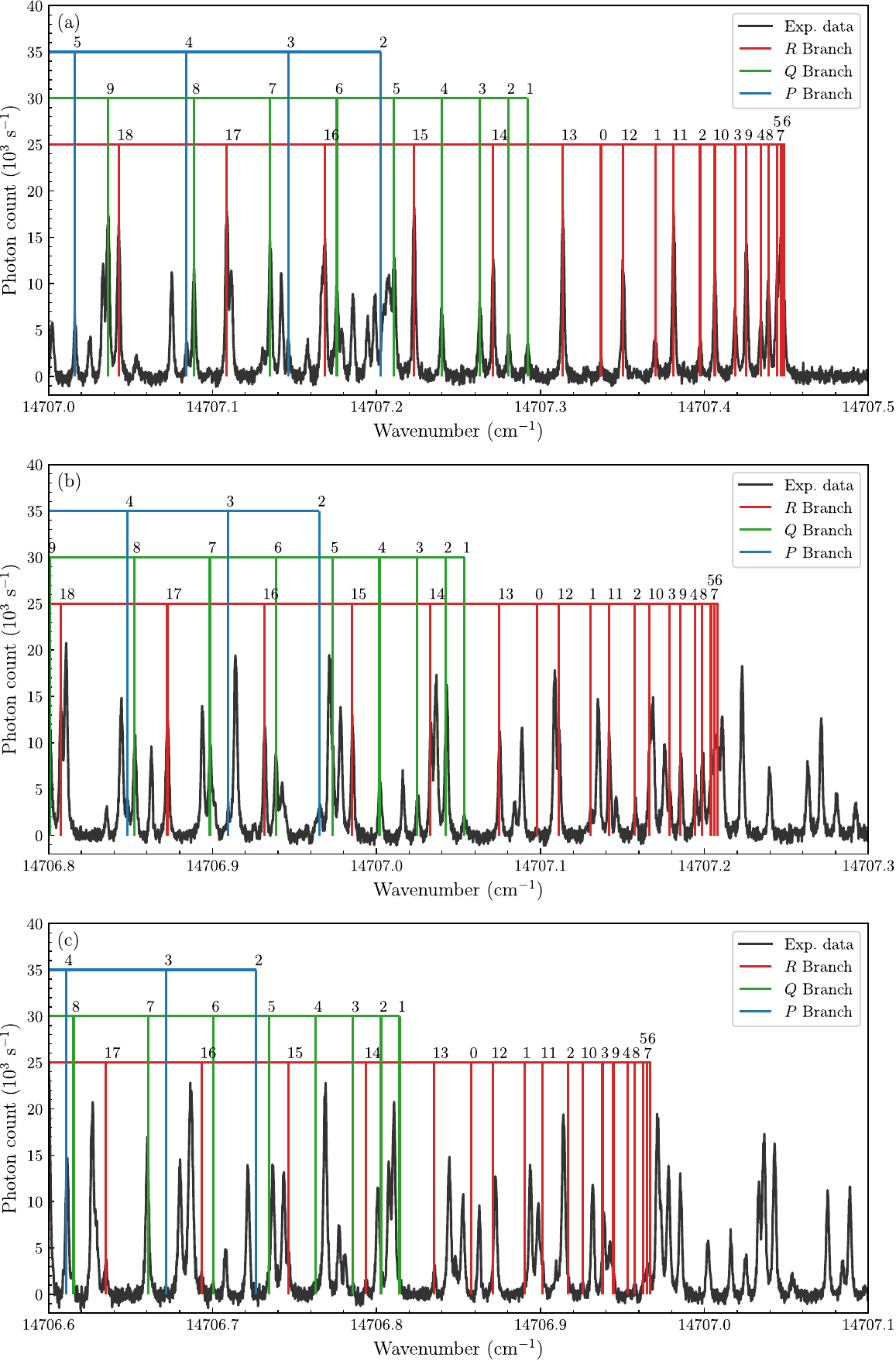}
\caption{Fluorescence spectra recorded in a spectral range of $0.5~\mathrm{cm^{-1}}$ below the band head of the transition $X^1\Sigma_g^+(v''=0) \rightarrow B^1\Pi_u(v'=1)$ for the isotopologues (a) $^{85}\mathrm{Rb}_2$, (b) $^{85}\mathrm{Rb}^{87}\mathrm{Rb}$, (c) $^{87}\mathrm{Rb}_2$. The $R$, $Q$, and $P$ rotational branches are indicated in red, green, and blue, respectively. Line assignments are presented separately for each isotopologue to avoid excessive overlap of vertical markers. Lines not assigned in a given panel originate from transitions of other isotopologues. Therefore, all observed spectral lines are ultimately assigned.}
\label{fig:4}
\end{figure}

Subsequently, the same assignment procedure was applied to the other two isotopologues $^{85}\mathrm{Rb}^{87}\mathrm{Rb}$ and $^{87}\mathrm{Rb}_2$. Figure~\ref{fig:4} presents the spectra recorded $0.5~\mathrm{cm}^{-1}$ below the band head for  $^{85}\mathrm{Rb}_2$,  $^{85}\mathrm{Rb}^{87}\mathrm{Rb}$, and  $^{87}\mathrm{Rb}_2$. The total number of assigned transitions for each vibrational band and isotopologue are summarized in Table~\ref{tab:1}. Unassigned transitions mainly result from weak signal intensities relative to the noise level and partially unresolved rotational structures. As expected, the assignment efficiency decreases with the natural abundance of isotopologues and is lowest for $^{87}\mathrm{Rb}_2$. For the $v''=1 \rightarrow v'=1$ vibrational band, no transitions of $^{87}\mathrm{Rb}_2$ could be identified. The frequencies of all assigned rotational transitions are provided in the Supplementary Material \cite{supp}. Table~\ref{tab:2} summarizes the total number of rotational states of the excited electronic state $B^1\Pi_u$ accessed through the assigned transitions for each isotopologue. The complete assignment of the rotational states was achieved for $^{85}\mathrm{Rb}_2$ and $^{85}\mathrm{Rb}^{87}\mathrm{Rb}$, while more than two-thirds of the expected states were identified for $^{87}\mathrm{Rb}_2$, reflecting its lower natural abundance and consequently weaker spectral signal.

\begin{table}[htbp]
\caption{Total number of assigned rotational transitions for each vibrational band and isotopologue considered in this study. The numbers in parentheses indicate the number of rotational states expected to be excited from the fitted Dunham coefficients of the $X^1\Sigma_g^+$ and $B^1\Pi_u$, from Refs. \cite{Seto2000} and \cite{Amiot1997}, respectively.}
\footnotesize
\centering
\setlength{\tabcolsep}{3.5pt}
\begin{tabular}{|c|c|c|c|c|c|c|c|c|c|c|}
\hline
\multirow{2}{*}{$v''$} & \multirow{2}{*}{$v'$} & \multicolumn{3}{|c|}{$^{85}\mathrm{Rb}_2$} & \multicolumn{3}{|c|}{$^{85}\mathrm{Rb}^{87}\mathrm{Rb}$} & \multicolumn{3}{|c|}{$^{87}\mathrm{Rb}_2$} \\
\cline{3-11}
& & $P$ & $Q$ & $R$ & $P$ & $Q$ & $R$ & $P$ & $Q$ & $R$ \\
\hline
0 & 1 & 40(41) & 48(48) & 55(56) & 39(40) & 44(47) & 50(56) & 22(40) & 29(47) & 24(55)\\
\hline
0 & 2 & 38(41) & 48(48) & 50(56) & 38(39) & 43(46) & 48(54) & 22(38) & 37(45) & 34(53)\\
\hline
1 & 1 & 31(41) & 46(48) & 50(56) & 30(41) & 45(48) & 47(53) & - & -& - \\
\hline
\end{tabular}
\label{tab:1}
\end{table}

\begin{table}[htbp]
\caption{Total number of rotational states of the excited electronic state $B^1\Pi_u$ accessed through the transitions listed in Table \ref{tab:1}. The numbers in parentheses indicate the number of rotational states expected to be excited as in Table \ref{tab:1}.}
\footnotesize
\centering
\setlength{\tabcolsep}{3.5pt}
\begin{tabular}{|c|c|c|c|}
\hline
& $^{85}\mathrm{Rb}_2$ & $^{85}\mathrm{Rb}^{87}\mathrm{Rb}$ & $^{87}\mathrm{Rb}_2$\\
\hline 
$v'=1$ & 56(56) & 56(56) & 37(55)\\
\hline
$v'=2$ & 56(56) & 54(54) & 47(53)\\
\hline
\end{tabular}
\label{tab:2}
\end{table}

\section{Determination of Molecular Parameters from the Present Data}
\label{sec4}

After assigning the spectral lines for all isotopologues and all investigated vibrational bands, the Dunham and $\Lambda$-coupling coefficients were determined through a simultaneous fit of 958 transitions involving the $R$, $Q$, and $P$ branches of all isotopologues using Eq.~(\ref{eq:4}). The sign $+$ in Eq.~(\ref{eq:4}) was used for the $P$ and $R$ branches, while the  sign $-$ was used for the $Q$ branch. This procedure was carried out using a least-squares fitting approach implemented with the SciPy library in Python \cite{Virtanen2020}.

%As mentioned previously, the Dunham coefficients of the ground $\Sigma$ state are well established \cite{Seto2000}; therefore, they were fixed during the fitting procedure. Fig.~\ref{fig:5}(a) shows the measured energy of the $v'' = 0 \rightarrow v' = 1$ $R$ (red squares), $Q$ (green triangles) and $P$ (blue dots) branch transitions of $^{85}\mathrm{Rb}_2$ as a function of $J''$. The best fitting curves are also shown for the $R$ (red line), $Q$ (green line) and $P$ (blue line) branches. The residuals are presented in Fig.~\ref{fig:5}(b), indicating excellent agreement between the fitting and the experiment. Similar results are obtained for the other isotopologs and vibrational bands, with a standard deviation of $5.0 \times 10^{-4}~\mathrm{cm}^{-1}$ considering all the measured lines. 

\begin{figure}[H]
\centering
\includegraphics[width=\textwidth]{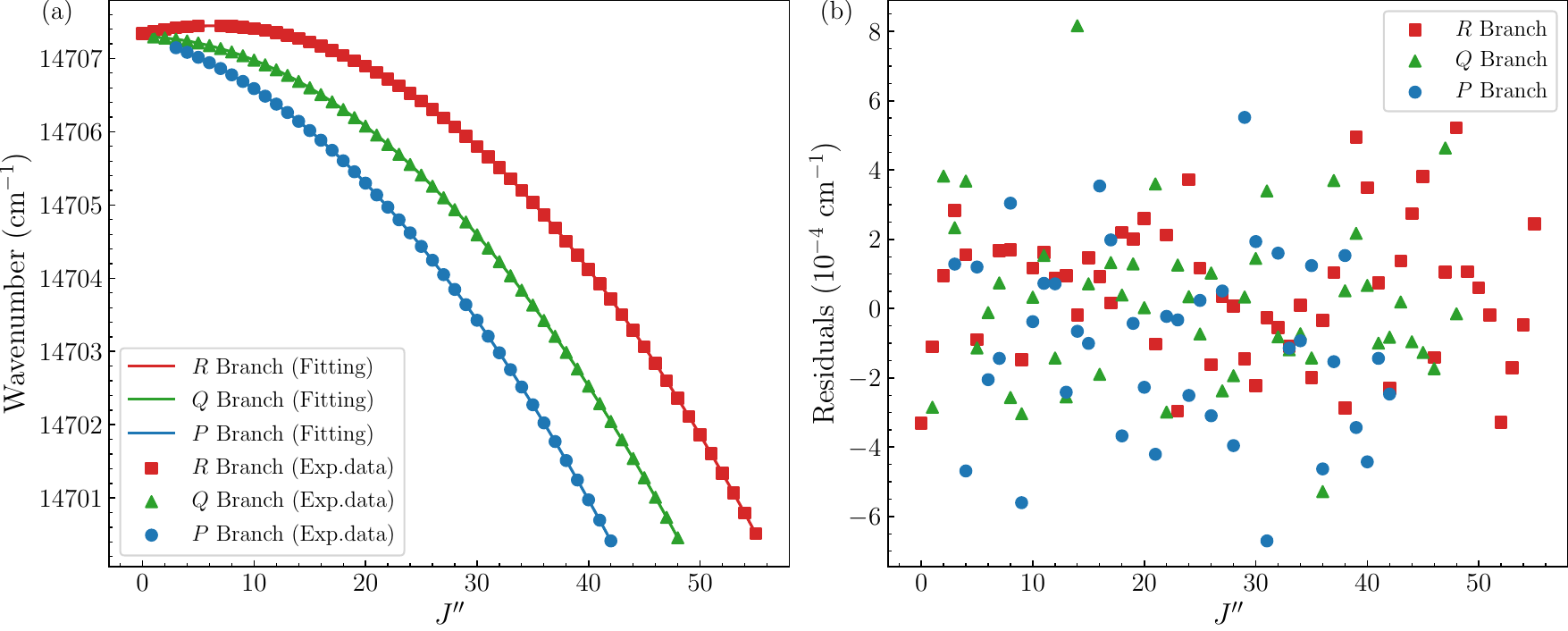}
\caption{(a) Simultaneous least-squares fit to the rotational transitions in the $R$ (red line), $Q$ (green line), and $P$ (blue line) branches of the $v'' = 0 \rightarrow v' = 1$ vibrational band of $^{85}\mathrm{Rb}_2$. The corresponding experimental data are shown as red squares ($R$ branch), green triangles ($Q$ branch), and blue circles ($P$ branch). (b) Residuals of the fit, demonstrating excellent agreement between the fitted model and the experimental data.}
\label{fig:5}
\end{figure}

As mentioned previously, the Dunham coefficients of the ground state $X^1\Sigma_g^+$ are well established \cite{Seto2000}; therefore, they were kept fixed during the fitting procedure. Figure~\ref{fig:5}(a) displays wavenumbers of the $R$, $Q$ and $P$ lines of the vibrational band $v''=0 \rightarrow v'=1$ of $^{85}\mathrm{Rb}_2$ as a function of $J''$, with the corresponding best-fit curves. The residuals between the experimental and calculated transition energies are presented in Fig.~\ref{fig:5}(b), producing a standard deviation of $5.0\times10^{-4}~\mathrm{cm}^{-1}$, demonstrating the excellent agreement between the model and the experimental data. Equivalent figures for the other isotopologues and vibrational bands are provided in the Supplementary Material \cite{supp}.
 
Since the lines corresponding to the $v''=1\rightarrow v'=1$ band probe the same excited states as the $v''=0\rightarrow v'=1$ transitions but with a lower accuracy due to a reduced signal-to-noise ratio, they were excluded from the fit. This reduced the number of fitted transitions to 709 and resulted in an improved standard deviation of $2.8\times10^{-4}~\mathrm{cm}^{-1}$. Therefore, only the $v''=0\rightarrow v'=\{1,2\}$ vibrational bands were considered for the remainder of this work.

Table~\ref{tab:3} presents the Dunham coefficients and $\Lambda$-doubling constants obtained from the present fit, together with the results reported in \cite{Amiot1997} and \cite{drozdova2014}. In the present analysis, only the terms that were found to significantly improve the fit were retained, resulting in a reduced set of parameters compared to the previous work. For consistency, only the coefficients common to both analyzes are reported in the table.

\begin{table}[htbp]
\caption{$^{85}\mathrm{Rb}_2$ $B^1\Pi_u$ Dunham coefficients and $\Lambda$-doubling constants ($p_0$) determined in this work, compared with the corresponding values reported by C. Amiot and J. Vergès \cite{Amiot1997} and N. Drozdova \textit{et al.} \cite{drozdova2014}. The centrifugal distortion correction to the $\Lambda$-doubling constant, $q_0$, was not significant and was therefore omitted from the fit. Uncertainties represent one standard deviation.}
\centering
\setlength{\arrayrulewidth}{1pt}
\setlength{\tabcolsep}{1.0pt}
\scriptsize

\begin{tabular}{|c|c|c|c|}

\hline
 & This work $(\mathrm{cm}^{-1})$ & Reference \cite{Amiot1997}$(~\mathrm{cm}^{-1})$ & Reference \cite{drozdova2014}$(~\mathrm{cm}^{-1})$\\
\hline
$Y_{0,0}'$& $14666.3018\pm0.0131$  & $14665.6482\pm0.0072$ & $14665.4540\pm0.0045$\\
\hline
$Y_{1,0}'$ & $46.3903\pm0.0133$  & $47.0999\pm0.0032$ & $47.35499\pm0.00052$\\
\hline
$Y_{3,0}'$ & $(1.47578\pm0.03099)\times10^{-1}$ & $-(0.4029\pm0.0040)\times10^{-1}$ & $-(3.447\pm0.100)\times10^{-4}$
\\
\hline
$Y_{4,0}'$ & $-(4.11211\pm0.07302)\times10^{-2}$ & $-(0.70244\pm0.01278)\times10^{-2}$ & $-(2.50\pm0.42)\times10^{-6}$\\
\hline
$Y_{0,1}'$ & $(1.955801\pm0.000018)\times10^{-2}$ &  $(1.954631\pm0.000145)\times10^{-2}$ & $(1.956201\pm0.000140)\times10^{-2}$\\
\hline
$Y_{1,1}'$ & $-(5.9766\pm0.0088)\times10^{-5}$ & $-(5.5739\pm0.0339)\times10^{-5}$ & $-(5.9041\pm0.0055)\times10^{-5}$\\
\hline
$Y_{0,2}'$ & $-(1.3315\pm0.0070)\times10^{-8}$ & $-(1.2309\pm0.0093)\times10^{-8}$ & $-(1.377\pm0.0011)\times10^{-8}$\\
\hline
$Y_{1,2}'$ & $-(1.79\pm0.34)\times10^{-10}$  & $-(4.4986\pm0.2452)\times10^{-10}$ & $-(0.715\pm0.026)\times10^{-10}$\\
\hline
$p_0$      & $(4.27\pm0.21)\times10^{-7}$ & -------- & --------\\
\hline
\multicolumn{4}{|l|}{709 transitions $\sigma = 2.8\times10^{-4}~\mathrm{cm}^{-1}$  } \\
\hline

\end{tabular}
\label{tab:3}
\end{table}

Our pure vibrational Dunham coefficients ($Y'_{k,0}$) show noticeable differences compared to those reported in Refs.~\cite{Amiot1997, drozdova2014}, together with larger uncertainties for some coefficients. This behavior is expected, since our determination is based on the observation of rovibronic transitions corresponding only to $B^1\Pi_u~(v'=1,2)$ vibrational states of the three isotopologues, which provides less sensitivity to the purely vibrational coefficients. In contrast, the rotational and ro-vibrational Dunham coefficients benefit from the large number of measured rotational transitions and the high spectral resolution of the present data.

%\textcolor{red}{WE SHOULD BE CAUTIOUS HERE }: For the first-order pure rotational Dunham coefficient, $Y'_{0,1}$, also known as the rotational constant $B_e$, we obtained a value that differs from that reported in Ref.~\cite{Amiot1997}. Our result presents one order of magnitude better precision, reflecting the higher spectral resolution of our experimental data. 
%\textcolor{red}{INDEED: Your formulation is clear, but it is somewhat too assertive in three respects: “Differs from that reported” may suggest a significant discrepancy, whereas the two values are likely compatible within their respective uncertainties. “One order of magnitude better precision” is slightly too strong, indeed the improvement is closer to a factor of 8 (as indicated by your uncertainty values). It would be more appropriate to refer to a reduction in the uncertainty or to an improvement in precision approaching one order of magnitude. “Reflecting the higher spectral resolution” implies a direct causal relationship. A more cautious phrasing would be to state that the improvement is consistent with or can be attributed to the higher spectral resolution of the experimental data}

For the first-order pure rotational Dunham coefficient, $Y'_{0,1}$, corresponding to the rotational constant $B_e$, our value is in good agreement with that reported in Refs.~\cite{Amiot1997, drozdova2014}. The present determination achieves a significantly smaller uncertainty, approaching an improvement of one order of magnitude, which can be attributed to the higher spectral resolution of our experimental data.

A similar improvement is obtained for the first-order rovibrational coupling Dunham coefficient, $Y'_{1,1}$. Our value is consistent with that reported in ~\cite{Amiot1997,drozdova2014}, while providing an uncertainty reduced by a factor of about four. For the Dunham coefficients associated with centrifugal distortion and their corresponding first-order coupling terms, our values lie outside the uncertainty ranges reported in ~\cite{Amiot1997, drozdova2014}. Nevertheless, the uncertainties obtained for these coefficients remain of the same order of magnitude as those reported previously.

Finally, we emphasize that the $\Lambda$-doubling constant, $p_0$, was determined in the present work with uncertainty below $5\%$, as shown in the last line of Table~\ref{tab:3}. To our knowledge, this is the first determination of the $\Lambda$-doubling constant within a global Dunham analysis of the $B^1\Pi_u$ state. No significant contribution from a second-order $\Lambda$-doubling term $q_0$, was required to reproduce the experimental data. Caldwell \textit{et al.}~\cite{Caldwell1980} also reported a $\Lambda$-doubling constant derived from separate fits of the $Q$ branch and the $P$/$R$ branches. However, because their analysis relies on a different fitting procedure and parameterization, a direct numerical comparison with the present value of $p_0$ is not straightforward.

Since the determination of the pure vibrational Dunham coefficients, $Y_{k,0}$, was not sufficiently constrained by the present data set, which is limited to only two excited vibrational levels, we adopted an alternative fitting strategy. In this approach, the pure vibrational contribution was incorporated through independent band-origin terms, $\nu^{(i)}(v'',v')$, for each vibrational band and isotopologue (Eq.~\ref{eq:8}), while common rotational and rovibrational coupling Dunham coefficients were retained for all three isotopologues, as described in Eq.~\ref{eq:2}. The quality of the fit remains comparable to that obtained in the previous analysis, and the resulting parameters are reported in Table~\ref{tab:4}. In this table, the listed values of the band origins from~\cite{Amiot1997, drozdova2014} were not obtained from a direct fit of the band origins, but were reconstructed from the reported pure vibrational Dunham coefficients. The excellent agreement between our present band origins and those reconstructed ones demonstrates the consistency of the two approaches, despite the different treatment of the vibrational contribution. The present band origins exhibit uncertainties that are more than two orders of magnitude smaller, reflecting the higher spectral resolution of the present measurements. The rotational and rovibrational coupling Dunham coefficients obtained from this alternative approach agree, within their respective uncertainties, with those determined in the previous fit, confirming the robustness of our fitted rotational spectroscopic coefficients. Compared to our previous determination reported in Table~\ref{tab:3}, the $\Lambda$-doubling constant, $p_0$, is found to be unchanged within its uncertainty when adopting this alternative fitting strategy, indicating that its determination is not affected by the different treatment of the pure vibrational contribution.

\begin{table}[htbp]
\caption{Best-fit parameters obtained from the alternative fitting procedure, in which the pure vibrational contribution is represented by independent band-origin terms, $\nu^{(i)}(v'',v')$, for each vibrational band and isotopologue. The table includes the first- and second-order pure rotational Dunham coefficients, the first-order rovibrational coupling Dunham coefficients, and the $\Lambda$-doubling constant. For comparison, the table also includes the band origin frequencies calculated using the $B^1\Pi_u$ Dunham coefficients reported in \cite{Amiot1997} and \cite{drozdova2014}, both referenced to the $X^1\Sigma_g^+$ potential of Ref. \cite{Seto2000}. The corresponding Dunham coefficients, as well as the rovibrational coupling Dunham coefficients reported in these references, are also listed.}
\centering
\setlength{\arrayrulewidth}{1pt}
\setlength{\tabcolsep}{1.0pt}
\scriptsize

\begin{tabular}{|c|c|c|c|}
\hline
 & This work $(\mathrm{cm}^{-1})$ & Reference \cite{Amiot1997}$~(\mathrm{cm}^{-1})$ & Reference \cite{drozdova2014}$~(\mathrm{cm}^{-1})$ \\
\hline
$\nu^{(1)}(0,1)$ & $14707.31751\pm0.00003$ & $14707.33246\pm0.00893$ & $14707.31772\pm0.00461$ \\
\hline
$\nu^{(2)}(0,1)$ & $14707.07882\pm0.00004$ & $14707.09313\pm0.00891$ & $14707.07807\pm0.00461$ \\
\hline
$\nu^{(3)}(0,1)$ & $14706.83879\pm0.00004$ & $14706.85239\pm0.00889$ & $14706.83700\pm0.00461$ \\
\hline
$\nu^{(1)}(0,2)$ & $14754.11753\pm0.00004$ & $14754.11796\pm0.01372$ &  $14754.12094\pm0.00478$ \\
\hline
$\nu^{(2)}(0,2)$ & $14753.61237\pm0.00004$ & $14753.61190\pm0.01360$ & $14753.61479\pm0.00477$ \\
\hline
$\nu^{(3)}(0,2)$ & $14753.10420\pm0.00004$ & $14753.10282\pm0.01347$ & $14753.10563\pm0.00477$ \\
\hline
$Y_{0,1}'$ & $(1.955797\pm0.000018)\times10^{-2}$ &  $(1.954631\pm0.000145)\times10^{-2}$ & $(1.956201\pm0.000140)\times10^{-2}$\\
\hline
$Y_{1,1}'$ & $-(5.9763\pm0.0088)\times10^{-5}$ & $-(5.5739\pm0.0339)\times10^{-5}$ & $-(5.9041\pm0.0055)\times10^{-5}$\\
\hline
$Y_{0,2}'$ & $-(1.3273\pm0.0070)\times10^{-8}$ & $-(1.2309\pm0.0093)\times10^{-8}$ & $-(1.377\pm0.0011)\times10^{-8}$\\
\hline
$Y_{1,2}'$ & $-(1.81\pm0.34)\times10^{-10}$  & $-(4.4986\pm0.2452)\times10^{-10}$ & $-(0.715\pm0.026)\times10^{-10}$\\
\hline
$p_0$      & $(4.28\pm0.21)\times10^{-7}$ & -------- & \\
\hline
\multicolumn{4}{|l|}{709 transitions $\sigma = 2.8\times10^{-4}~\mathrm{cm}^{-1}$  }\\
\hline
\end{tabular}
\label{tab:4}
\end{table}

\section{Global Determination of Molecular Parameters Using Present and Literature Data}
\label{sec5}

C. Amiot and J. Vergès determined 40 Dunham coefficients by fitting 1186 rovibrational levels belonging to two $\mathrm{Rb}_2$ isotopologues \cite{Amiot1997}. Their analysis provided an extensive description of the excited state $B^1\Pi_u$ over a broad range of vibrational and rotational quantum numbers. However, the available data were mainly concentrated at relatively high values of $v'$ and $J'$, and the experimental spectra were recorded with a resolution of 0.01~$\mathrm{cm}^{-1}$. The present measurements considerably extend and improve upon this previous spectroscopic investigation. Although restricted to the $v'=1$ and $v'=2$ vibrational levels of the excited $B^1\Pi_u$ state, our data provide substantially enhanced spectral resolution and a dense set of low-$J'$ rotational transitions for all three $\mathrm{Rb}_2$ isotopologues. They therefore offer complementary information to the dataset of \cite{Amiot1997}, where the wider vibrational coverage is essential to determine the vibrational dependence of the Dunham coefficients. To evaluate the impact of the present measurements on the spectroscopic model of \cite{Amiot1997}, a global least-squares analysis \cite{LeRoy1998, LeRoy1999, Bouloufa2000} was performed by combining the newly measured transitions with the previously available rovibrational levels.

\begin{figure}[H]
\centering
\includegraphics[width=0.75\textwidth]{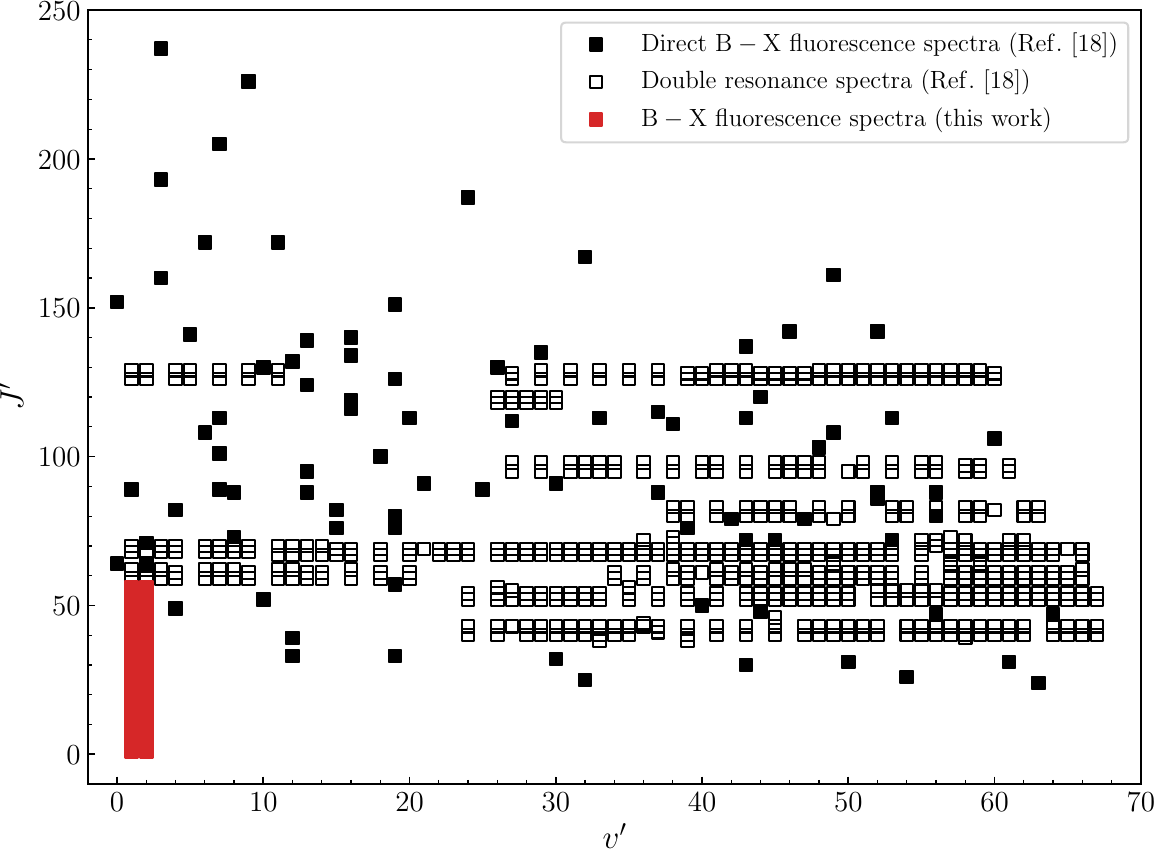}
\caption{Rotational and vibrational quantum numbers of the $B^1\Pi_u$ rovibrational levels included in the global fitting analysis. Black squares denote the states reported by Amiot and Vergès \cite{Amiot1997} and reconstructed from their Dunham coefficients. Filled black squares represent the 87 levels directly excited through $B-X$ transitions, whereas open black squares correspond to the 524 levels populated via double-resonance excitation followed by fluorescence decay. Red squares indicate the 112 excited levels experimentally accessed in the present work for $^{85}\mathrm{Rb}_2$.}
\label{fig:6}
\end{figure}

Since the transition frequencies and rovibrational states reported in Ref. \cite{Amiot1997} are not publicly available, the excited-state rovibrational levels included in their analysis were extracted from Fig. 5 of that work. The corresponding energy levels were then reconstructed using the Dunham coefficients reported therein. It should be noted that the states populated through rotational relaxation are not explicitly identified in Fig.~5; therefore, their assignment was retained as given in the original analysis. Figure \ref{fig:6} summarizes all rovibrational states included in the present global fit. In total, 611 excited-state levels were extracted from Ref. \cite{Amiot1997}. They include 87 states issued by direct $X \rightarrow B$ fluorescence transitions (shown as filled black squares) and 524 states corresponding to double-resonance transitions followed by spontaneous decay (shown as open black squares). The complete list of reconstructed levels is provided in the Supplementary Material \cite{supp}. The excited states rovibrational levels measured in the present work are displayed as red squares in Fig. \ref{fig:6}; for $^{85}\mathrm{Rb}_2$, they correspond to 112 additional levels.

\begin{figure}[H]
\centering
\includegraphics[width=0.75\textwidth]{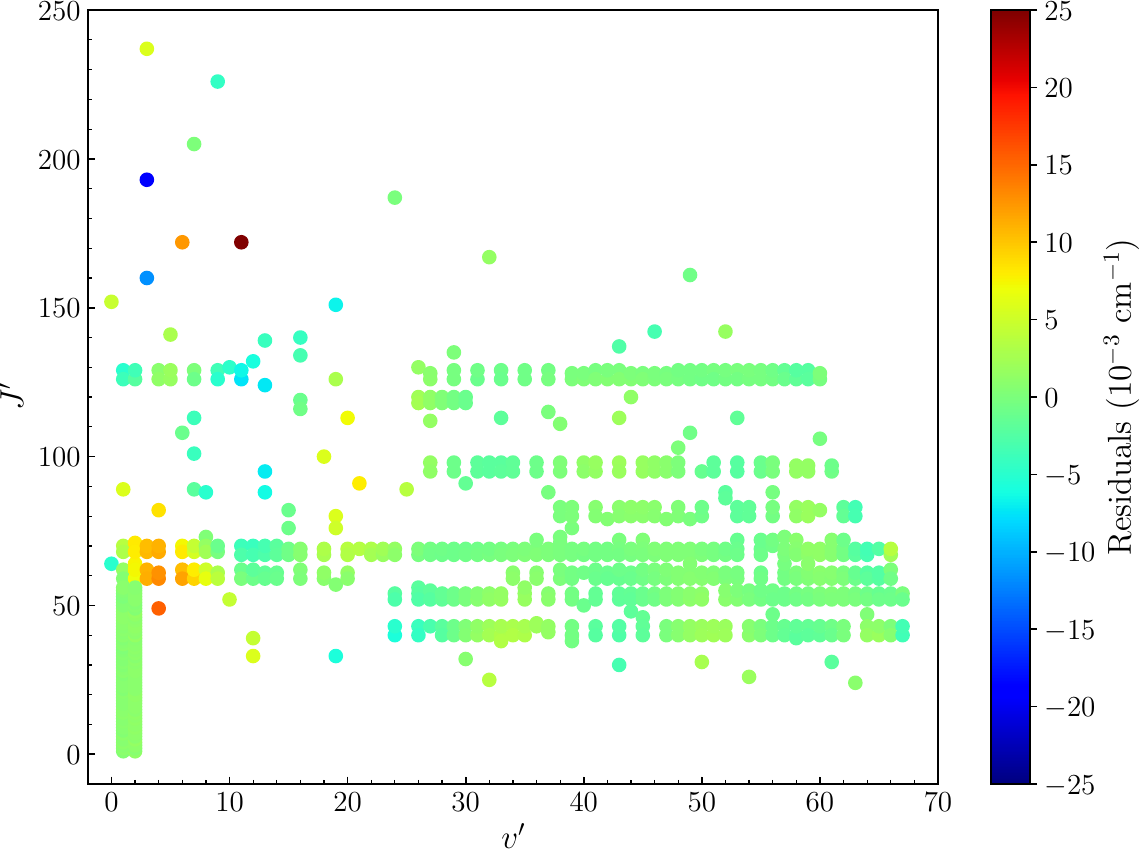}
\caption{Residuals obtained from the global fitting procedure as a function of the ro-vibrational quantum numbers. The figure includes the residuals for the experimental data reported by Amiot and Vergès \cite{Amiot1997}, as well as those measured in the present work. The standard deviation is  $\sigma = 6.6\times10^{-4}~\mathrm{cm}^{-1}$.}
\label{fig:7}
\end{figure}

\begin{figure}[h]
\centering
\includegraphics[width=1\textwidth]{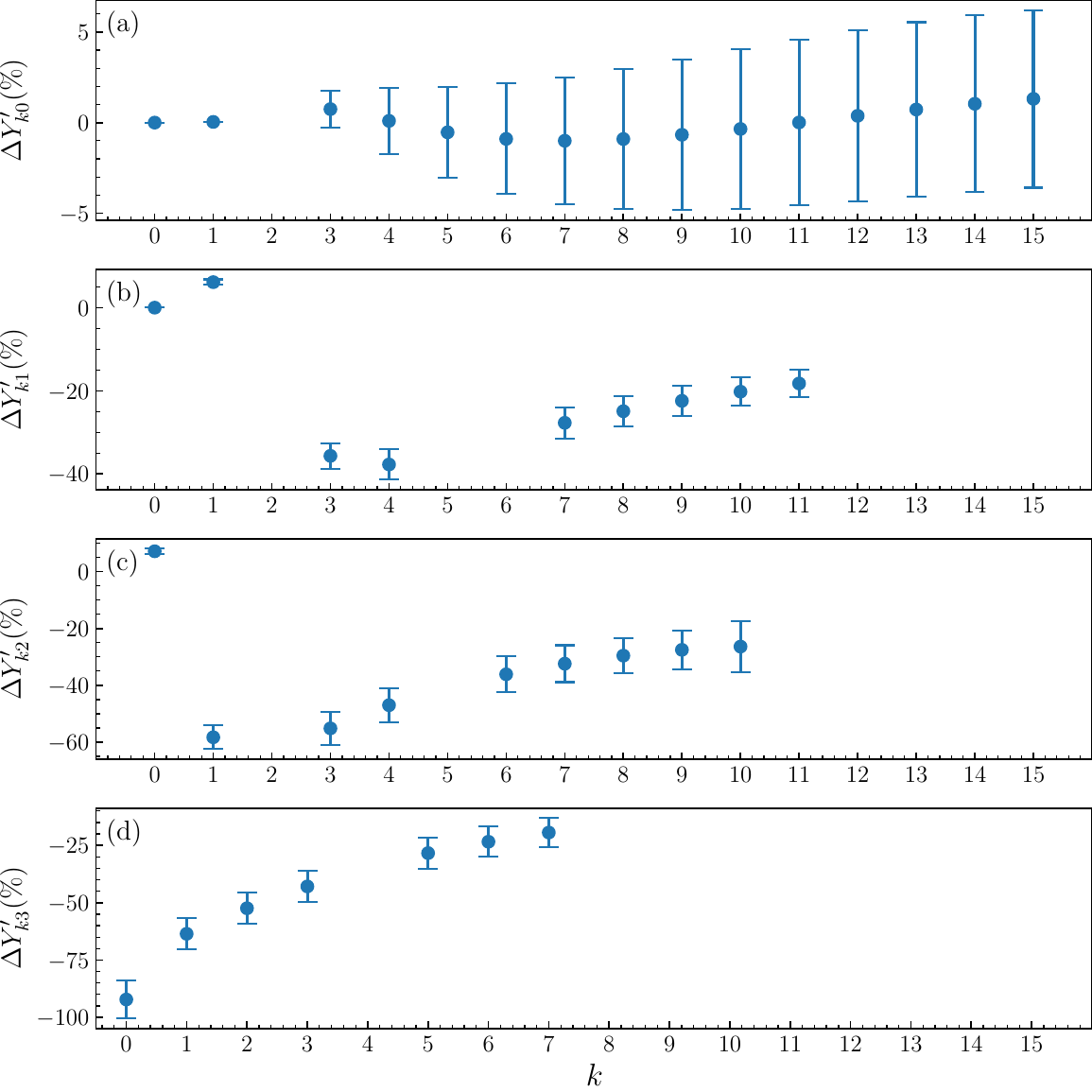}
%\caption{Relative percentage differences between the Dunham parameters $(\Delta Y'_{kl})$ determined from the global fitting procedure developed in the present work and those reported by Amiot and Vergès \cite{Amiot1997}. Panel (a) presents the pure vibrational parameters $(\Delta Y'_{k0})$, while panels (b), (c), and (d) display the first-order $(\Delta Y'_{k1})$, second-order $(\Delta Y'_{k2})$, and third-order $(\Delta Y'_{k3})$ rotational parameters, respectively, together with their corresponding ro-vibrational coupling contributions.}

\caption{Relative percentage differences, $\Delta Y'_{k,l}(\%)$, between the Dunham parameters determined from the global fitting procedure developed in the present work and those reported by C. Amiot and J. Vergès (1). Panel (a) corresponds to the pure vibrational parameters ($\Delta Y'_{k,0}$), while panels (b), (c), and (d) display the first-, second-, and third-order ro-vibrational parameters ($\Delta Y'_{k,1}$, $\Delta Y'_{k,2}$, and $\Delta Y'_{k,3}$), respectively. The error bars represent the propagated standard uncertainties of the relative differences obtained from the uncertainties of the two compared Dunham parameters. The horizontal zero line corresponds to perfect agreement between the two determinations.}
\label{fig:8}
\end{figure}

For global fitting analysis, a least-squares fitting procedure was implemented using the SciPy Python library \cite{Virtanen2020}. The residuals were weighted according to the standard deviations obtained from independent fits performed separately for the present dataset and for the dataset of Amiot and Vergès. The corresponding standard deviations used as weighting factors are $\sigma_\mathrm{P}=2.8 \times 10^{-4}~\mathrm{cm}^{-1}$ and $\sigma_\mathrm{A}=4.9 \times 10^{-3}~\mathrm{cm}^{-1}$, respectively. The residuals resulting from the global fit for the $^{85}\mathrm{Rb}_2$ isotopologue are presented as a function of the vibrational and rotational quantum numbers of the corresponding states in Fig.~\ref{fig:7}. For the present dataset, the residuals remain of the same order of magnitude as those obtained in the independent fits described in Section~\ref{sec4}, indicating that the inclusion of the Amiot and Vergès data does not degrade the quality of the fit to the new measurements. In contrast, the residuals associated with the Amiot and Vergès dataset are largest for the lowest vibrational levels, which overlap with the vibrational region investigated in the present work, and progressively decrease with increasing vibrational quantum number. The global analysis yields a standard deviation of $\sigma = 6.6 \times 10^{-4}~\mathrm{cm}^{-1}$. Table~\ref{tab:5} presents the Dunham coefficients obtained from the present global fit, denoted as $Y'^{(P)}_{k,l}$, together with the corresponding coefficients reported by Amiot and Vergès, $Y'^{(A)}_{k,l}$, and the relative differences between the two sets of coefficients, defined as ($\Delta Y'_{k,l} = (Y'^{(P)}_{k,l}-Y'^{(A)}_{k,l})\times100/Y'^{(A)}_{k,l}$). Figure~\ref{fig:8} displays the relative differences $\Delta Y'_{k,l}$. The results are organized as follows: Fig.~\ref{fig:8}(a) presents the purely vibrational parameters, while Figs.~\ref{fig:8}(b), \ref{fig:8}(c), and \ref{fig:8}(d) show the
first-, second-, and third-order purely rotational parameters, respectively, together with their corresponding rovibrational coupling coefficients.

The results of the global fitting procedure are discussed in the following. The impact of the present measurements on the Dunham coefficients is examined by comparing the fitted parameters with those reported in ~\cite{Amiot1997}. First, the global fitting procedure does not lead to significant changes in the purely vibrational parameters within their respective uncertainties (Fig.~\ref{fig:8}(a)). This behavior is consistent with the limited vibrational coverage of the present dataset, which is restricted to only two low-lying vibrational levels. Consequently, the vibrational information provided by the larger dataset of Amiot and Vergès \cite{Amiot1997} remains essentially unchanged in the combined analysis.

The rotational terms exhibit a different behavior. The first-order purely rotational coefficients ($Y'_{01}$) obtained from the global fit show small but measurable positive deviations exceeding their estimated uncertainties, while the first-order rovibrational coupling coefficients ($Y'_{11}$) display more pronounced positive shifts (Fig.~\ref{fig:8}(b)). The influence of the present measurements is therefore mainly reflected in the first-order rovibrational coupling terms. In contrast, the higher-order rovibrational coupling coefficients exhibit compensating negative deviations, allowing the combined model to maintain consistency with the higher-lying states reported in Ref. \cite{Amiot1997}, as shown in Fig. \ref{fig:8}(b).

A similar behavior is observed for the second-order purely rotational parameter ($Y'_{02}$), which presents a significant positive deviation beyond its uncertainty (Fig.~\ref{fig:8}(c)). This change is compensated by negative adjustments in the higher-order rovibrational coupling coefficients.

The third-order purely rotational parameter ($Y'_{03}$) obtained from the global fit is not statistically significant (Fig. \ref{fig:8}(d)), indicating the limited sensitivity of the available data to third-order rotational effects. This may also result from the limited number of levels extracted from Ref. \cite{Amiot1997}, for which only 611 states could be included in the global analysis. The additional states observed through rotational relaxation, which are not explicitly identified in the original work, may have involved higher rotational quantum numbers and could therefore have provided stronger constraints on the third-order rotational parameters.

Finally, the rotational and rovibrational coupling coefficients ($Y'_{01}$, $Y'_{11}$, $Y'_{02}$, and $Y'_{12}$) obtained from the global fit are in closer agreement with the coefficients determined in the previous section where only the present experimental data were analyzed (see Tables~\ref{tab:3} and \ref{tab:4}). This confirms the significant improvement induced by the higher spectral resolution and the extended low-$J'$ rotational coverage of the present dataset.

\begin{table}[htbp]
\caption{Dunham coefficients obtained from the present global fitting procedure and those reported in ref~\cite{Amiot1997}. The uncertainties correspond to one standard deviation. We also list the relative percentage difference $\Delta Y'_{k,l}(\%)$ with its uncertainty obtained by propagation of the standard deviations of the two coefficients. A value of $\Delta Y'_{k,l}$ statistically compatible with zero indicates agreement between the two determinations. Therefore, a small value of $\Delta Y'_{k,l}$ compared with its uncertainty (e.g., $0.7 \pm 1.0\%$) represents consistency rather than poor accuracy.}
\centering
\setlength{\arrayrulewidth}{1pt}
\setlength{\tabcolsep}{1.0pt}
\scriptsize
\begin{tabular}{|c|c|c|c|}
\hline
 & This work $(\mathrm{cm}^{-1})$ & Reference \cite{Amiot1997}$(~\mathrm{cm}^{-1})$ & $\Delta Y'_{k,l}~(\%)$\\
\hline
$Y_{0,0}'$ & $14665.607973\pm0.000480$  & $14665.648165915\pm0.007274$ & $-(2.7\pm0.5)\times10^{-4}$\\
\hline
$Y_{1,0}'$ & $47.117864093\pm0.000409$  & $47.099863700374\pm0.003284$ & $(3.8\pm0.7)\times10^{-3}$\\
\hline
$Y_{3,0}'$ & $-(4.0590374070\pm0.006668)\times10^{-2}$ & $-(4.0289857593501\pm0.03978)\times10^{-2}$ & $0.7\pm1.0$\\
\hline
$Y_{4,0}'$ & $(7.0312489989\pm0.0160818)\times10^{-3}$ & $-(7.0244588598746\pm0.1278)\times10^{-3}$ & $0.1\pm1.8$\\
\hline
$Y_{5,0}'$ & $-(8.0465168092254\pm0.00173978) \times10^{-4}$ & $-(8.0034472162\pm0.2016) \times10^{-4}$ & $-(0.5\pm2.5)$\\
\hline
$Y_{6,0}'$ & $ (6.2731596399\pm0.0107398) \times10^{-5}$ & $ (6.3300244404850\pm0.1946) \times10^{-5}$ & $-(0.9\pm3.0)$\\
\hline
$Y_{7,0}'$ & $-(3.5071215113\pm0.00413451) \times10^{-6}$ & $-(3.5426555481958\pm0.1253) \times10^{-6}$ & $-(1.0\pm3.5)$\\
\hline
$Y_{8,0}'$ & $(1.4264090569\pm0.00104127)\times10^{-7}$ & $(1.4394621467149\pm0.05623)\times10^{-7}$ & $-(0.9\pm3.9)$\\
\hline
$Y_{9,0}'$ & $-(4.2569645765   \pm0.00179236)\times10^{-9}$ & $-(4.2857310363   \pm0.1798)\times10^{-9}$ & $-(0.7\pm4.2)$\\
\hline
$Y_{10,0}'$ & $(9.3092562298\pm0.00201312)\times10^{-11}$ & $(9.3420622544911\pm0.4129)\times10^{-11}$ & $-(0.4\pm4.4)$\\
\hline
$Y_{11,0}'$ & $-(1.4732710138\pm0.0000956864)\times10^{-12}$ & $-(1.4731431286453\pm0.06757)\times10^{-12}$ & $0.009\pm4.487$\\
\hline
$Y_{12,0}'$ & $(1.6413136564\pm0.000461597)\times10^{-14}$ & $(1.6351890914594\pm0.07696)\times10^{-14}$ & $0.4\pm4.7$\\
\hline
$Y_{13,0}'$ & $-(1.2203253803\pm0.000780991)\times10^{-16}$ & $-(1.2115694135393\pm0.05797)\times10^{-16}$ & $0.7\pm4.8$\\
\hline
$Y_{14,0}'$ & $(5.4345084931\pm0.00587867)\times10^{-19}$ & $(5.3786847177919\pm0.2596)\times10^{-19}$ & $1.0\pm4.9$\\
\hline
$Y_{15,0}'$ & $-(1.0962854880\pm0.00173806)\times10^{-21}$ & $-(1.0820933028853\pm0.005235)\times10^{-21}$ & $1.3\pm4.9$\\
\hline
$Y_{0,1}'$ & $(1.9557146932\pm0.0000204958)\times10^{-2}$ &  $(1.9546311449987\pm0.0001451)\times10^{-2}$ & $(5.5\pm0.8)\times10^{-2}$\\
\hline
$Y_{1,1}'$ & $-(5.9197223900\pm0.00926903)\times10^{-5}$ & $-(5.5739469089063\pm0.03393)\times10^{-5}$ & $6.2\pm0.7$\\
\hline
$Y_{3,1}'$ & $-(5.7451188101\pm0.191100)\times10^{-8}$ & $-(8.9311547169872\pm0.3076)\times10^{-8}$ & $-(35.7\pm3.1)$\\
\hline
$Y_{4,1}'$ & $(2.4155052283\pm0.108663)\times10^{-9}$ & $(3.8797117982920\pm0.1464)\times10^{-9}$ & $-(37.7\pm3.6)$\\
\hline
$Y_{7,1}'$ & $-(2.3290566125\pm0.0972864)\times10^{-13}$ & $-(3.2214686948299\pm0.1008)\times10^{-13}$ & $-(27.7\pm3.8)$\\
\hline
$Y_{8,1}'$ & $(1.2534710344\pm0.0493738)\times10^{-14}$ & $(1.6691104774836\pm0.04893)\times10^{-14}$ & $-(24.9\pm3.7)$\\
\hline
$Y_{9,1}'$ & $-(2.7807347543\pm0.102655)\times10^{-16}$ & $-(3.5837368344049\pm0.09851)\times10^{-16}$ & $-(22.4\pm3.6)$\\
\hline
$Y_{10,1}'$ & $(2.8998473079\pm0.0999902)\times10^{-18}$ & $(3.6336040089392\pm0.09378)\times10^{-18}$ & $-(20.2\pm3.4)$\\
\hline
$Y_{11,1}'$ & $-(1.1759600952\pm0.0377711)\times10^{-20}$ & $-(1.4378783601854\pm0.03488)\times10^{-20}$ & $-(18.2\pm3.3)$\\
\hline
$Y_{0,2}'$ & $-(1.3187493548\pm0.00470671)\times10^{-8}$ & $-(1.2309155951338\pm0.009313)\times10^{-8}$ & $7.1\pm0.9$\\
\hline
$Y_{1,2}'$ & $-(1.8772961784\pm0.154730)\times10^{-10}$  & $-(4.4985553497972\pm0.2452)\times10^{-10}$ & $-(58.3\pm4.1)$\\
\hline
$Y_{3,2}'$ & $(2.2910097392\pm0.264440)\times10^{-12}$ & $(5.1058441725387\pm0.2857)\times10^{-12}$ & $-(55.1\pm5.8)$\\
\hline
$Y_{4,2}'$ & $-(1.7933942755\pm0.178651)\times10^{-13}$ & $-(3.3823912599307\pm0.1725)\times10^{-13}$ & $-(47.0\pm5.9)$\\
\hline
$Y_{6,2}'$ & $(4.3304851168\pm0.373728)\times10^{-16}$ & $(6.7792933481940\pm0.3113)\times10^{-16}$ & $-(36.1\pm6.2)$\\
\hline
$Y_{7,2}'$ & $-(1.8129654399\pm0.153232)\times10^{-17}$ & $-(2.6816958298683\pm0.1213)\times10^{-17}$ & $-(32.4\pm6.5)$\\
\hline
$Y_{8,2}'$ & $(3.2657373187\pm0.281604)\times10^{-19}$ & $(4.6358076838237\pm0.02134)\times10^{-19}$ & $-(29.6\pm6.1)$\\
\hline
$Y_{9,2}'$ & $-(2.7542609973\pm0.255633)\times10^{-21}$ & $-(3.7999575943155\pm0.01862)\times10^{-21}$ & $-(27.5\pm6.7)$\\
\hline
$Y_{10,2}'$ & $(0.87099837609\pm0.0937968)\times10^{-23}$ & $(1.1829835347919\pm0.06566)\times10^{-23}$ & $-(26.4\pm8.9)$\\
\hline
$Y_{0,3}'$ & $-(1.8877571235\pm1.99342)\times10^{-15}$ & $-(2.4295526203753\pm0.2149)\times10^{-14}$ & $-(92.2\pm8.2)$\\
\hline
$Y_{1,3}'$ & $(5.1444719346\pm0.899027)\times10^{-15}$ & $(1.4120364618997\pm0.08686)\times10^{-14}$ & $-(63.6\pm6.7)$\\
\hline
$Y_{2,3}'$ & $-(8.4490808270\pm0.113514)\times10^{-16}$ & $-(1.7758789232999\pm0.09706)\times10^{-15}$ & $-(52.4\pm6.9)$\\
\hline
$Y_{3,3}'$ & $(4.2706037714\pm0.475741)\times10^{-17}$ & $(7.4793753147554\pm0.3667)\times10^{-17}$ & $-(42.9\pm6.9)$\\
\hline
$Y_{5,3}'$ & $-(5.6317983168\pm0.480374)\times10^{-20}$ & $-(7.8612905880856\pm0.3164)\times10^{-20}$ & $-(28.4\pm6.8)$\\
\hline
$Y_{6,3}'$ & $(1.4640703791\pm0.113418)\times10^{-21}$ & $(1.9114011654233\pm0.07097)\times10^{-21}$ & $-(23.4\pm6.6)$\\
\hline
$Y_{7,3}'$ & $-(1.1337811380\pm0.0795694)\times10^{-23}$ & $-(1.4065893519321\pm0.04804)\times10^{-23}$ & $-(19.4\pm6.3)$\\
\hline
\multicolumn{4}{|l|}{917 states $\sigma = 6.6\times10^{-4}~\mathrm{cm}^{-1}$  }\\
\hline
\end{tabular}
\label{tab:5}
\end{table}

\section{Conclusion}
\label{sec6}

In summary, this work significantly advances the spectroscopic understanding of the $\mathrm{Rb}_2$ molecule by performing high-resolution laser spectroscopy on a supersonic molecular beam. By resolving rovibrational transitions for three distinct isotopologues ($^{85}\mathrm{Rb}_2$, $^{85}\mathrm{Rb}^{87}\mathrm{Rb}$, and $^{87}\mathrm{Rb}_2$) with superior spectral resolution, we have not only refined earlier datasets from Caldwell \textit{et al.} \cite{Caldwell1980}, C. Amiot and J. Vergès \cite{Amiot1997}, and Drozdova \textit{et al.} \cite{drozdova2014}, but also provided new measurements of the $\Lambda$-doubling in the state. Although an attempt to improve the spectroscopic description of the $B^1\Pi_u$ state was previously reported in the PhD thesis of Drozdova \cite{Drozdova2012}, not all of the corresponding experimental data were included in the subsequent publication \cite{drozdova2014}, leaving some questions open regarding the spectroscopic description of the lowest vibrational levels. The subsequent global fit presented here, which integrates our high-precision data with those of C. Amiot and J. Vergès \cite{Amiot1997}, therefore provides a more complete and accurate description of the excited-state structure. These results may be useful for other research areas \cite{Hu2021,Liu2021,PASHOV2022,PhysRevA.105.032823,PhysRevA.109.012805,Alejandro2023,Lefran2025}.

%\od{(OD: I suggest that the work of Drozdova could be invoked here, by saying that data are available in the thesis as an attempt to improve the B state, but not all the data is reported in the corresponding paper, thus leaving some questions open)}
%In summary, this work significantly advances the spectroscopic understanding of the $\mathrm{Rb}_2$ molecule by performing high-resolution laser spectroscopy on a supersonic molecular beam. By resolving rovibrational transitions for three distinct isotopologues ($^{85}\mathrm{Rb}_2$,  $^{85}\mathrm{Rb}^{87}\mathrm{Rb}$, and  $^{87}\mathrm{Rb}_2$) with superior spectral resolution, we have not only refined earlier datasets from Caldwell \textit{et al.} \cite{Caldwell1980} and C. Amiot and J. Vergès \cite{Amiot1997} but also provided new measurements of the $\Lambda$-doubling in the state. The subsequent global fitting, which integrates our high-precision data with those of C. Amiot and J. Vergès \cite{Amiot1997}, yields a more complete and accurate description of the excited-state structure. These results may be useful for other research areas \cite{Hu2021,Liu2021,PASHOV2022,PhysRevA.105.032823,PhysRevA.109.012805,Alejandro2023,Lefran2025}.

\section{Acknowledgements}
This work is supported by Grants 2018/06835-0, 2022/16904-5, 2023/06732-5, and 2021/04107-0 from São Paulo Research Foundation (FAPESP), FA9550-23-1-0666 from the US Air Force Office of Scientific Research, 305257/2022-6 from CNPq, and ANR-21-CE30-0060-01 (COCOTRAMOS project) from the Agence Nationale de la Recherche. L. G. M. thanks the support of Université Paris-Saclay for generous support as a visiting professor.

\bibliographystyle{unsrt}  % or plain, apalike, etc.
\bibliography{references}

\end{document}